\documentclass[12pt]{article}
\usepackage{cjp}
\newcommand {\ee}{\end{equation}}
\newcommand {\bea}{\begin{eqnarray}}
\newcommand {\eea}{\end{eqnarray}}
\newcommand {\nn}{\nonumber \\}
\newcommand {\Tr}{{\rm Tr\,}}

\newcommand {\e}{{\rm e}}

\newcommand {\m}{\mu}

\newcommand {\pl}{\partial}

\newcommand {\al}{\alpha}

\newcommand {\ga}{\gamma}
\newcommand {\Ga}{\Gamma}

\newcommand {\om}{\omega}
\newcommand {\Om}{\Omega}

\newcommand {\na}{\nabla}
\newcommand {\del}  {\delta}

\newcommand {\half}{ {\frac{1}{2}} }

\newcommand {\Dvec}{{\hat D}}   

\newcommand {\Dhat}{{\hat D}}

\newcommand {\psibar}{{\bar \psi}}

\newcommand {\ra} {\rightarrow}

\newcommand {\pr}   {{\quad .}}
\newcommand {\com}  {{\quad ,}}
\newcommand {\q}    {\quad}

\newcommand {\nl}    {\newline}

\newcommand {\NP}   {Nucl.Phys.}
\newcommand {\PL}   {Phys.Lett.}
\newcommand {\PR}   {Phys.Rev.}
\newcommand {\PRL}   {Phys.Rev.Lett.}

\newcommand {\JMP}  {Jour.Math.Phys.}

\def\overleftrightarrow#1{\vbox{\ialign{##\crcr
 $\leftrightarrow$\crcr\noalign{\kern-1pt\nointerlineskip}
 $\hfil\displaystyle{#1}\hfil$\crcr}}}

\newcommand {\gago}  {\gamma_5}
\newcommand {\Ktil}  {{\tilde K}}

\newcommand {\Kbar}  {{\bar K}}
\newcommand {\GfMp}  {G^{5M}_+}

\newcommand {\GfMm}  {G^{5M}_-}
\def\Aslash{{}\hbox{\hskip2pt\vtop
 {\baselineskip23pt\hbox{}\vskip-24pt\hbox{/}}
 \hskip-11.5pt $A$}}
\def\kslash{{}\hbox{\hskip2pt\vtop
 {\baselineskip23pt\hbox{}\vskip-24pt\hbox{/}}
 \hskip-8.5pt $k$}}
\def\dbslash{{}\hbox{\hskip2pt\vtop
 {\baselineskip23pt\hbox{}\vskip-24pt\hbox{$\backslash$}}
 \hskip-11.5pt $\partial$}}
\newcommand {\dslash} { {\not\partial}}  

\def\Ktilbslash{{}\hbox{\hskip2pt\vtop
 {\baselineskip23pt\hbox{}\vskip-24pt\hbox{$\backslash$}}
 \hskip-11.5pt ${\tilde K}$}}

\def\Kbarbslash{{}\hbox{\hskip2pt\vtop
 {\baselineskip23pt\hbox{}\vskip-24pt\hbox{$\backslash$}}
 \hskip-11.5pt ${\bar K}$}}

\begin{document}
\title{Heat-Kernel Approach to the Overlap Formalism 
\footnote{
Talk at CHIRAL'99,Taipei,Sep.13-18,1999;\ Univ. of Shizuoka
preprint, US-99-09
}
       }
\author{Shoichi Ichinose}
\address{Department of Physics, University of Shizuoka \\
Yada 52-1, Shizuoka 422-8526, Japan. }
\date{\today}
\maketitle

\begin{abstract}
We present a new regularization method, for 
d dim (Euclidean) quantum field theories in the continuum formalism, 
based on the domain wall configuration in (1+d) dim space-time.
It is inspired by the recent progress in the 
chiral fermions on lattice. 
The wall "height" is given by $1/M$, where $M$ is 
a regularization mass parameter and appears as a
(1+d) dim Dirac fermion mass. The present approach gives 
a {\it thermodynamic view} to the domain wall or the overlap formalism
in the lattice field theory. 
We will show qualitative correspondence between the present 
continuum results and those of lattice.
The extra dimension 
is regarded as the (inverse) {\it temperature} $t$.
The domains are defined by the {\it directions} of
the "system evolvement", not by
the sign of $M$ as in the original overlap formalism.
We take 
the 4 dim QED 
and 2 dim chiral gauge theory as examples.
Especially the consistent and covariant
anomalies are correctly obtained. 
\end{abstract}


$\mbox{}$\nl
{\large I.\ Introduction}\q 
Regularizing quantum theories respecting the chirality has been a long-lasting problem
both in the discrete and in the continuum field theories. 
The difficulty originates
from the fact that the chiral symmetry is a symmetry 
strongly bound to the space-time dimension 
and is related to the discrete symmetry of parity
and to the global features of the space-time topology.
Non-continuous property is usually difficult to regularize. 
Ordinary regularizations, such as the dimensional regularization, 
often hinder controlling the chirality.
The symmetry should be compared
with others such as the gauge symmetry of the internal space
and the Lorentz symmetry of the space-time.
In the lattice field theory, the difficulty appears as the doubling
problem of fermions\cite{Wil75} (see a text, say, \cite{Creu83})
and as the Nielsen-Ninomiya no-go theorem
\cite{NN81}. 
The recent very attractive progress in the lattice chiral fermion
tells us the domain wall configuration in one dimension higher
space(-time) serves as a good regularization,
at least, as far as vector theories are concerned
\cite{Kap92,Jan92}. 
It was formulated as the overlap formalism\cite{NN94,NN95}
and was further examined by \cite{DS95PL,DS95NP,DS97PL}.
The corresponding lattice models were analyzed by 
\cite{Sha93,CH94,Creu95,FS95,Vra98}.
The numerical data also look to support its validity\cite{Col99}.
Most recently 
the overlap Dirac operator by Neuberger\cite{Neub98},
which satisfies Ginsparg-Wilson relation\cite{GW82}, and
L{\"{u}}scher's chiral symmetry on lattice\cite{Lus98}
makes the present direction more and more attractive.

The overlap formalism has been 
newly formulated using the heat-kernel\cite{SI98}. 
The heat-kernel formalism is most efficiently expressed in the
coordinate space\cite{II98}, and which enables us to do comparison
with the lattice formalism. We will often
compare the present results with those obtained by 
the lattice domain wall approach.
Through the analysis we expect to clarify the essence of
the regularization mechanism more transparently than on lattice. 
We see some advanced points over the  
ordinary regularizations in the continuum field theories. 
The main goal is to develop a new
feasible regularization, in the continuum formalism,  
which is compatible with the chiral symmetry.

The present formalism is based on three key points:
\begin{enumerate} 
\item
We utilize the characteristic relation of heat and temperature, 
that is, heat propagates
from the high temperature to the low temperature 
(the second law of thermodynamics). 
In the system which obey the heat equation, there exists
a {\it fixed direction} in the system evolvement. 
We regard the heat equation
for the spinor system, after the Wick rotation,  
as the Dirac equation in one dimension higher space-time. And
we consider the (1+4 dim classical) configuration 
which has a fixed direction in time.
This setting is suitable for regularizing the dynamics
(in 4 dim Euclidean space) with control of the chirality. 
\item 
Anti-commutativity between the system operator $\Dhat$
and the chiral matrix, that is, $\gago\Dhat+\Dhat\gago=0$
plays the crucial role 
to separate the whole configuration
into two parts (we will call them "(+)-domain" and "(-)-domain") 
which are related by the sign change of the "time"-axis. 
This is contrasted with
the original formulation of the overlap where
the difference of two vacua, one is constructed from
the (+) sign regularization (1+4 dim fermion) mass and the other from
the (-) sign,
distinguishes the two domains.
\item
Taking the small momentum region compared to the regularization
mass scale $M$ regularize the ultra-violet divergences 
and, at the same time, controls the chirality.
\end{enumerate}

$\mbox{}$\nl
{\large II.\ Domain Wall Regularization}\q
Let us analyze
the 4 dim massless Euclidean QED in the domain wall approach.
First we express the effective action 
$\ln\,Z[A]=\Tr\ln\Dvec$ 
in terms of 
$\Dvec=i\ga_\m(\pl_\m+i\e A_\m)$ itself (not its square).
Formally we have
\begin{eqnarray}
\Tr\ln\Dvec=-\Tr\int_0^\infty\frac{\e^{-t\Dvec}}{t}dt
=-\int_0^\infty\frac{dt}{t}\Tr
[\half(1+i\ga_5) \e^{+it\ga_5\Dvec}+
 \half(1-i\ga_5) \e^{-it\ga_5\Dvec}].
\label{DW2}
\end{eqnarray}
Because the eigenvalues of $\Dhat$ are both negative and positive,
the $t$-integral above is divergent.
We clearly need regularization to make it meaningful.
We should notice here that the final equality above
relies only on the following properties of $\Dhat$ and $\gago$:\ 
$\ga_5\Dvec+\Dvec\ga_5=0\ ,\ (\gago)^2=1\ .$
Note that, in the final expression of (\ref{DW2}),
the signs of the eigenvalues of $\Dhat$ become less
important
for the $t$-integral convergence. This is because
the {\it exponential} operator $\e^{-t\Dhat}$ is replaced
by the {\it oscillating} operators $\e^{\pm it\gago\Dhat}$
due to the above properties
\footnote{
This statement is a disguise at the present stage. It is correct
after the {\it Wick rotations} for $t$ after (\ref{DW4b}).
}
. 
Here we introduce a regularization parameter $M$,
which is most characteristic in this approach.
$M$ is taken to be positive for simplicity:\ $M>0$. 
\begin{eqnarray}
\ln\,Z=\Tr\ln\Dvec=-\lim_{M\ra 0}\int_0^{\infty}\frac{dt}{t}
\half(1-i\frac{\pl}{t\,\pl M})\Tr (\GfMp(x,y;t)+\GfMm(x,y;t))\com
\label{DW4}
\end{eqnarray}
where
$G^{5M}_\pm (x,y;t)\equiv <x|\exp\{\pm it\ga_5(\Dvec+iM)\}|y>$.
\footnote{
The case of $M=0$ reduces to (\ref{DW2}) where the relation 
$\ga_5\Dvec+\Dvec\ga_5=0$ is essential.
In this sense we can consider this regularization procedure
is related with the generalization of the above anti-commutation relation
in some $M$-dependent way.
It looks to correspond to
a kind of Ginsparg-Wilson relation\cite{GW82}.
}
$M$ can be regarded as the "source" for $\gago$.
Through this procedure we can treat $\gago$ within the new
heat kernels $G^{5M}_\pm$.
From its usage above, the limit $M\ra 0$ should be taken in the
following way before $t$-integral:
\begin{eqnarray}
Mt\ll 1\pr
\label{DW4b}
\end{eqnarray}
Very interestingly, the above heat-kernels satisfy 
the same 1+4 dim
Minkowski Dirac equation after the following {\it Wick rotations} for $t$\ :\ 
$
(i\dbslash-M)G^{5M}_\pm =i\e \Aslash G^{5M}_\pm \ ,\ 
(X^a)=(\mp it,x^\m)\ ,
$                    
where $\Aslash\equiv\ga_\m A_\m(x)\ ,\ 
\dbslash\equiv\Ga^a\frac{\pl}{\pl X^a}$. (
$\m=1,2,3,4;\ a=0,1,2,3,4$) 
Note here that
the sign of the Wick-rotation is different for $\GfMp$ and for $\GfMm$.
$\GfMp$ and $\GfMm$ turn out to
correspond to (+)-domain and (-)-domain, respectively, 
in the original formulation \cite{NN94,NN95}
and we also call them in the same way.

Both $\GfMp$ and $\GfMm$ are obtained in the same form $G^5_M$
specified by ($G_0,S$).
\begin{eqnarray}
G^5_M(X,Y)=G_0(X,Y)+\int d^5Z\,S(X,Z)i\e\Aslash(z)G^5_M(Z,Y)\com
\label{DW6}
\end{eqnarray}
where 
$G_0(X,Y)$ is the free solution and $S(X,Z)$ is the propagator:\ 
$(i\dbslash-M)G_0(X,Y)=0\ ,\ 
(i\dbslash-M)S(X,Y)=\del^5(X-Y)$.
There are four choices of the propagator. 
(See Fig.1 of \cite{SI99}.) 
From them we make three solutions
(see Sec.III).
They are obtained by 
some combinations of the positive and negative
energy free solutions:
\begin{eqnarray}
G^p_0(X,Y)\equiv -i\int\frac{d^4k}{(2\pi)^4}\Om_+(k)\e^{-i\Ktil(X-Y)}
\equiv \int\frac{d^4k}{(2\pi)^4}G^p_0(k)\e^{-ik(x-y)}\ ,\nn 
G^n_0(X,Y)\equiv -i\int\frac{d^4k}{(2\pi)^4}\Om_-(k)\e^{+i\Kbar(X-Y)}
\equiv \int\frac{d^4k}{(2\pi)^4}G^n_0(k)\e^{-ik(x-y)}\ ,
\label{DW8}
\end{eqnarray}
where 
$ \Om_+(k)\equiv (M+\Ktilbslash)/2E(k),\ 
G^p_0(k)\equiv -i\Om_+(k)\e^{-iE(k)(X^0-Y^0)},\ 
\Om_-(k)\equiv (M-\Kbarbslash)/2E(k),\ 
G^n_0(k)\equiv -i\Om_-(k)\e^{iE(k)(X^0-Y^0)},\ 
E(k)=\sqrt{k^2+M^2},\ (\Ktil^a)=(\Ktil^0=E(k),\ \Ktil^\m=-k^\m),\ 
(\Kbar^a)=(\Kbar^0=E(k),\Kbar^\m=k^\m).$ 
$k^\m$ is the momentum in the 4 dim Euclidean space.
$\Ktil$ and $\Kbar$ are on-shell
momenta($\Ktil^2=\Kbar^2=M^2$), which correspond to the positive
and negative energy states respectively.
We can rewrite $\Om_\pm(k)$ as
\begin{eqnarray}
\Om_+(k)=\half (\gago+\frac{M+i\kslash}{|M+i\kslash|})\com\q 
\Om_-(k)=\half (-\gago+\frac{M+i\kslash}{|M+i\kslash|}) 
\label{DW8.5}
\end{eqnarray}
where $|M+i\kslash|\equiv E(k)$. 
The factor 
$\frac{M+i\kslash}{|M+i\kslash|}$ can be regarded as a "phase" operator 
{\it depending on configuration}.
The expressions above
look similar to the overlap Dirac operator
\cite{Neub98} for the case of {\it no} Wilson term.
\footnote{
The overlap Dirac operator $D$ on lattice is\ 
$ S_F=a^4\sum_{x\in \mbox{all sites}}\psibar(x)D\psi(x),\ 
aD=1+\gago\frac{H}{\sqrt{H^2}},\ $\nl
$\gago H=\sum_\m \{ \half \ga_\m(\na_\m+\na_\m^*)
-\frac{a}{2}\na_\m^*\na_\m \}-M,$\ 
where $a$ is a lattice spacing. If we ignore the 
$\na_\m^*\na_\m$-term (Wilson term), the form is
quite similar to (\ref{DW8.5}).
}
In fact, $\Om_\pm(k)$ have {\it projective property} with their hermite
conjugate. The relations 
will
be efficiently used in anomaly calculations.

The final important stage is regularization of the (1-loop) 
ultraviolet divergences. 
Corresponding to the 1-loop quantum evaluation, the determinant (\ref{DW2})
finally involves one momentum($k^\m$)-integral (besides $t$-integral).
We will take the analytic continuation method in order to avoid
introducing further regularization parameters and to avoid breaking the
gauge invariance. 
It can be shown\cite{SI99} that the method is essentially equivalent
to restricting the integral region from $0\leq |k^\m| < \infty$ to
\begin{eqnarray}
\mbox{Chiral Condition}\ :\ 
0\leq |k^\m| \leq M\pr
\label{DW8a}
\end{eqnarray}
\footnote{
Instead of the analytic continuation, we can take the higher derivative
regularization. This corresponds to the Wilson term in lattice:\ 
$i\dslash -M\ \ra\ i\dslash\pm\frac{r}{M}\pl^2-M,$ 
$r$: "Wilson term" coefficient. In this case the unitarity problem,
rather than the gauge invariance, should be clarified. 
}
This looks similar to the usual Pauli-Villars procedure in the point of
ultra-violet regularization. $M$ plays the role of the momentum cut-off.
We should stress that this restriction condition (\ref{DW8a})
on the momentum integral, at the same time, controls the chirality
as explained in the following. ( This point is a distinguished property
of the domain wall regularization. ) We call (\ref{DW8a}) {\it chiral condition}.
Its {\it extreme} case:
\begin{eqnarray}
\mbox{Extreme Chiral Limit}\ :\ 
\frac{M}{|k^\m|}\ra \infty\com
\label{DW8aa}
\end{eqnarray}
implies the chirality selection:
\begin{eqnarray}
\mbox{for}\ \mbox{(+)-domain}\ (X^0=-it)\q,
iG^p_0(k)\ra  \frac{1+\gago}{2}\e^{-Mt}\ , \ 
iG^n_0(k)\ra  \frac{1-\gago}{2}\e^{+Mt}\ ;\nn
\mbox{for}\ \mbox{(-)-domain}\ (X^0=+it)\q,
iG^p_0(k)\ra  \frac{1+\gago}{2}\e^{+Mt}\ , \ 
iG^n_0(k)\ra  \frac{1-\gago}{2}\e^{-Mt}\ . 
\label{DW8b}
\end{eqnarray}
This result will be used for characterizing different configurations
with respect to the chirality.
We use (\ref{DW8a}) instead of (\ref{DW8aa}) in the concrete
calculation. (\ref{DW8aa}) is {\it too restrictive} to keep the dynamics. 
Loosening the extreme chiral limit (\ref{DW8aa}) to the chiral condition
(\ref{DW8a}) can be regarded as a part of the present regularization.
This situation looks similar to the introduction of the Wilson term, 
in the lattice formalism, in order
to break the chiral symmetry. 

Let us reexamine the condition (\ref{DW4b}). As read from
the above result, the domain is characterized by the exponential
damping behavior which
has the "width"$\sim 1/M$ around the
origin of the extra $t$-axis. (\ref{DW4b}) restricts the region
of $t$ as $t\ll 1/M$. This is for considering only the massless mode
as purely as possible.
In the lattice formalism, this corresponds to taking
the zero mode (surface state) limit, in order to avoid the doubling problem,
by introducing many "flavor" fermions (or adding an extra dimension)
and many bosonic Pauli-Villars fields to kill the heavy fermions contribution. 
Besides the extreme chiral limit (\ref{DW8aa}), we often consider 
, corresponding to (\ref{DW4b}), the following limit:\  
\begin{eqnarray}
M|X^0-Y^0|\ra +0\pr
\label{DW8bx}
\end{eqnarray}
This limit will be taken to characterize the
full solutions (\ref{DW6})
by their {\it boundary conditions}.

In the lattice numerical simulation, 
the best fit value of the regularization mass $M$ looks
restricted both from the below and from the above
depending on the simulation "environment"
\cite{BS97PR,Col99}.
\footnote{
In the lattice formalism 
the corresponding bound on $M$
has been known, from the requirement of no doublers, 
since the original works\cite{Kap92,NN94}.
}
($M\sim$a few Gev for the hadron simulation.)
The similar one occurs in the present regularization.
The "double" limits (\ref{DW4b}) and 
(\ref{DW8a}) or (\ref{DW8aa}) imply
\begin{eqnarray}
|k^\m|\ll M\ll \frac{1}{t}\q \mbox{or}\q
|k^\m|\leq M\ll \frac{1}{t} \pr
\label{DW14}
\end{eqnarray}
In the standpoint of the extra dimension, the limit $M\ll \frac{1}{t}$ 
($Mt\ra +0$) corresponds to
, combined with the condition on $|k^\m|/M$, 
taking the dimensional reduction from
1+4 dim to 4 dim (Domain wall picture of 4 dim space).
The relation (\ref{DW14}) is the most characteristic one of the 
present regularization. It should be compared with the usual heat-kernel
regularization where only
the limit $t\ra +0$ is taken and the ultraviolet regularization
is done by the simple subtraction of divergences. 
Eq.(\ref{DW14}) shows the delicacy in taking the limit in the
the present 1+4 dimensional
regularization scheme. 
It implies, in the lattice simulation, 
$M$ should be appropriately chosen depending
on the regularization scale (,say, lattice size) and the  
momentum-region of 4 dim fermions. 

$\mbox{}$\nl
{\large III.\ (+)-Domain and (-)-Domain ($G^{5M}_\pm$)}\q
{\large III-1.\ Feynman Path and Anti-Feynman Path}\q
First we consider 
the Feynman propagator:\ 
$S_F(X,Y)=\theta (X^0-Y^0)G^p_0(X,Y)+\theta (Y^0-X^0)G^n_0(X,Y)$.
It has both the retarded and advanced parts.
Now we remind ourselves of the fact that there exists
a fixed direction in the system evolvement when the temperature
parameter works well.
Let us regard the extra axis, after the Wick-rotations,
as a temperature.  
Assuming the analogy holds here, we try to
adopt the following {\it directed} solution displayed in TABLE I.

\begin{tabular}{|c|c|c|}
\hline
                      &
   $\GfMp$(Retarded)  & 
   $\GfMm$(Advanced)                            \\
\hline
    $G_0$                  & 
    $G^p_0(X,Y)$ &
    $G^n_0(X,Y)$         \\
\hline
$S$ & 
$\theta (X^0-Y^0)G^p_0(X,Y)$    & 
  $ \theta (Y^0-X^0)G^n_0(X,Y)$   \\
\hline
 F.E.& 
$(i\dbslash-M)\GfMp =i\e \frac{1+\gago}{2}\Aslash \GfMp +O(\frac{1}{M})$   &
$(i\dbslash-M)\GfMm =i\e \frac{1-\gago}{2}\Aslash \GfMm +O(\frac{1}{M})$   \\
\hline
\multicolumn{3}{c}{TABLE I.\ \ (+)-domain and (-)-domain in 
the Feynman path solution.  }\\
\end{tabular}
This is chosen in such a way that the $t$-integral converges.
Because we have "divided" a full solution into two chiral parts in order to introduce
a {\it fixed direction} in the system evolvement, 
$G^{5M}_\pm$ defined in TABLE I 
do not satisfy 
$(i\dbslash-M)G^5_{M}=i\e \Aslash G^5_{M}$ but satisfy
, in the extreme chiral limit, the field eq. of the {\it chiral} QED:\ 
$\Dhat_\pm\equiv i(\dslash+ie\frac{1\pm\gago}{2}\Aslash)$\ . 
Taking the extreme chiral limit in the momentum spectrum, 
we can read off
the domain wall structure near the origin of the extra axis.
(See Fig.2 of \cite{SI99}.)
The full solutions ${G^{5M}_\pm}$ (\ref{DW6}) of TABLE I satisfy 
the boundary condition :
\begin{eqnarray}
i(\GfMp(X,Y)-\GfMm(X,Y))\ra \gago\del^4(x-y)\ \mbox{as}\ 
M|X^0-Y^0|\ra +0\ ,\nn
i(\GfMp(X,Y)+\GfMm(X,Y))\ra 
\int\frac{d^4k}{(2\pi)^4}\frac{M+i\kslash}{|M+i\kslash|}\e^{-ik(x-y)}
\ \mbox{as}\ M|X^0-Y^0|\ra +0\ .
\label{DW9d}
\end{eqnarray}
Taking into account
the boundary conditions above, we should take,
in the Adler-Bell-Jackiw and Weyl anomaly calculations, as
\begin{eqnarray}
\half\del_\al\,\ln\,J_{ABJ}=\lim_{M|X^0-Y^0|\ra +0}
\Tr\,i\al(x)i(\GfMp(X,Y)-\GfMm(X,Y))\com\nn
\half\del_\om\,\ln\,J_W=\lim_{M|X^0-Y^0|\ra +0}
\Tr\,\om(x)i\gago (\GfMp(X,Y)-\GfMm(X,Y))\pr
\label{DW10}
\end{eqnarray}

The meaning of the choice of Feynman path solution
(TABLE I)
is subtle (but interesting),
because the solution does not satisfy 
$(i\dbslash-M)G^5_{M}=i\e \Aslash G^5_{M}$. The clear separation
of right and left and its calculational simplicity fascinate
us to examine this solution. 

We can take the opposite choice
of $G^p_0$ and $G^n_0$ in TABLE I.
We call this case anti-Feynman path solution. 
The regularization using this solution turns out to
give the same result as the Feynman path solution. 
The different point is 
that, due to the presence of the exponentially growing
factor $\e^{+E(k)t}$, we must do calculation in
the $X^0$-coordinate.\cite{SI98}

$\mbox{}$\nl
{\large III-2.\ Symmetric Path}\q
Let us consider the symmetric pathes.
In this case we are led to take the solution displayed in TABLE II.

\begin{tabular}{|c|c|c|}
\hline
                      &
   $\GfMp$(Retarded)  & 
   $\GfMm$(Advanced)                            \\
\hline
    $G_0$                  & 
    $G^p_0(X,Y)-G^n_0(X,Y)$ &
    $G^n_0(X,Y)-G^p_0(X,Y)$         \\
\hline
$S$ & 
$\theta (X^0-Y^0)(G^p_0(X,Y)-G^n_0(X,Y))$    & 
  $ \theta (Y^0-X^0)(G^n_0(X,Y)-G^p_0(X,Y))$   \\
\hline
 F.E.& 
$(i\dbslash-M)\GfMp =i\e \Aslash \GfMp $   &
$(i\dbslash-M)\GfMm =i\e \Aslash \GfMm $   \\
\hline
\multicolumn{3}{c}{TABLE II.\ \ (+)-domain and (-)-domain in 
the symmetric path solution.  }\\
\end{tabular}
\nl
$G^{5M}_\pm$ satify the QED field equation properly.
Taking the extreme chiral limit $\frac{|k^\m|}{M}\ll 1$  
in the solution of TABLE II, we can read off
the symmetric wall structure 
(one wall at the origin and the other at the infinity,
see Fig.4 of \cite{SI99}).
The above solutions satisfy the following boundary condition:
\begin{eqnarray}
\frac{i}{2}(\GfMp(X,Y)-\GfMm(X,Y))\ra \gago\del^4(x-y)\ \mbox{as}\ 
M|X^0-Y^0|\ra +0\ .
\label{DW13bx}
\end{eqnarray}
In this case, the anomalies 
are regularized as 
\begin{eqnarray}
\del_\al\,\ln\,J_{ABJ}
=\lim_{M|X^0-Y^0|\ra +0}\Tr\,i^2\al(x)\{\GfMp(X,Y)-\GfMm(X,Y)\}
\com\nn
\del_\om\,\ln\,J_{W}=
\lim_{M|X^0-Y^0|\ra +0}\Tr\,i\om(x)\gago\{\GfMp(X,Y)-\GfMm(X,Y)\}
\pr
\label{DW13bb}
\end{eqnarray}

Both in (\ref{DW10}) and in (\ref{DW13bb}),
the anomalies are expressed by the "difference" between $\GfMp$ 
and $\GfMm$ contributions. 
This exactly corresponds to the "overlap" equation in the
original formalism. The "difference" in the effective action corresponds
to the "product" in the partition function between (+) part and (-) part,
that is the "overlap". This is the reason we name $G^{5M}_\pm$
as ($\pm$)-domains.

$\mbox{}$\nl
{\large IV.\ Anomaly Calculations}\q
Using the anomaly equations in Sec.III, 
explicit calculation has been done for 
2 dim QED, 4 dim QED and 2 dim chiral gauge theory.
As for the former two theories,
it is confirmed that the anomalies are correctly obtained
in the symmetric path solution, whereas $(1/2)^{d/2}$ times
of them for the Feynman path solution. As for the last theory,
the consistent and covariant anomalies are correctly obtained
by taking the chiral vertex for the former and the hermitian
vertex for the latter ( with the common choice of
the symmetric path solution). See \cite{SI98,SI99,SI99aei} for
details.

$\mbox{}$\nl
{\large IV.\ Discussion and Conclusion}\q
From the results of Sec.III,
we can imagine that the choice
of (anti-)Feynman path solution perturbatively defines the chiral version
of the original theory, for example, the chiral gauge theory. As far as
anomaly calculation is concerned, it holds true. In order to show the
statement definitely, we must clarify the following things. The "ordinary"
chiral symmetry appears only in the limit :\ $\frac{|k^\m|}{M}\ra +0$. 
But this limit can not be taken because it "freeze"
the dynamics and anomalies do {\it not} appear.  It seems we must introduce
some new "softened" version of the chiral symmetry which keeps the dynamics.
One standpoint taken in this paper is to replace 
$|k^\m|/M\ \ll 1$\ by $|k^\m|/M\ \leq 1$.
It breaks the "ordinary" chiral symmetry. It could, however, be
possible that this replacement can avoid the breaking
by changing (generalizing) the "ordinary" chiral symmetry. 
In this case, the new chiral Lagrangian has infinitely many higher-derivative
terms. We should explain this new "deformed" Lagrangian from some generalized
chiral symmetry. This situation looks similar to what
L\"{u}sher\cite{Lus98} did for the chiral lattice.

The chiral problem itself does not depend on the interaction.
It looks a kinematical problem in the quantization of fields.
How do we treat the different propagations of {\it free} solutions
depending on the
boundary conditions (with respect to the Wick-rotated time) 
is crucial to the problem.
In the standpoint of the operator formalism 
(the Fock-space formalism) 
it corresponds to how to treat the "delicate" structure
(due to the ambiguity of the fermion mass sign) 
of the vacuum of the free fermion theory.
The present paper insists the following prescription:\ 
First we go to 1+4 dim Minkowski space by the Wick-rotation
of the inverse temperature $t$, and take the "directed"
solution. 
The anomaly phenomena concretely reveal  
the chiral problem. The proposed prescription passes the anomaly test.


\end{document}